\documentstyle[pre,preprint,tighten,aps]{revtex}

\twocolumn

\begin{document}

\title{Stochastic Representation of Quantum Interactions
and Two-Lewel Systems}

\author{Yu.\,E.\,Kuzovlev\/\thanks{kuzovlev@kinetic.ac.donetsk.ua}}

\address{A.A.Galkin Physics and Technology Institute of NASU,
83114 Donetsk, Ukraine}

\date{10 September 2003}

\maketitle

\abstract{Stochastic representation for interaction of quantum
systems is formulated which allows to replace some of them
by equivalent but purely commutative random sources.
The formalism is applied to two-level systems interacting with
Gaussian thermal bath. Strong-coupling non-Marcovian effects
and besides long-living fluctuations in common susceptibility
of two systems subjected to the same bath are considered.

\,\,}

\,\,\,

\,\,\,

\,\,\,

{\bf 1. Stochastic representation.}

In quantum statistical physics and kinetics typically one deals
with tasks about interaction between some ``dynamical system''
(DS, microscopical or with only a few degrees of freedom)
and its macroscopic surroundings (``thermal bath'', or
``thermostat''). Usually, an interaction Hamiltonian has
bilinear structure (or can be reduced to that):
\begin{equation}
H_{int}=\sum_j B_j*D_j\,,
\end{equation}
where operators $D_j$ and $B_j$ act in different linear (Hilbert)
spaces $D$ and $B$ (native spaces of DS and thermostat,
respectively). In the Heisenberg picture, the operators
 $B_j(t)\equiv $  $\exp (iH_bt)B_j\exp (-iH_bt)\,$
(with $H_b $ being Hamiltonian of autonomous thermostat)
play like random perturbations with respect to DS.
But, in opposite to classical Langevin forces, they are
operator-valued non-commutative random sources and
therefore withdraw the DS evolution into the immense
direct product of the two spaces, $D\otimes B$.

Possibly, the task would become more simple,
if $B_j(t)$ were replaced by equivalent commutative
random sources, so that the DS evolution
would remain in $D\,$. The equivalence means exact
reproduction of results under interest,
first of all, exact simulation
of role of the thermostat temperature and
and effects of DS self-action mediated by thermostat,
in particular, dephasing and dissipation.

The variant of such a replacement was suggested in [1,5].
Let $H_d(t)$ be Hamiltonian of free DS. Introduce the
notation
\begin{equation}
H(B,t)=H_d(t)+\sum_j B_j(t)D_j
\end{equation}
This is joint Hamiltonian in the representation which is
Heisenbergian with respect to the thermostat.
Let some DS observables, $J_k $, are staying under
actual observation (measurement). Then statistical operator
of complete system ``DS plus thermostat'',  $R(t)\,$,
undergoes the equation
\begin{equation}
\dot{R}=v(t)J\circ R+i\left \{RH(B,t)-H(B,t)R \right \}\,,
\end{equation}
where $v(t)J\equiv$ $\sum v_k(t)J_k\,$,
with $v_k(t)$ being test functions of the observation
and $\circ $ the symbol of symmetric product (Jordan product).
At $v_k(t)=0\,$, this is the usual von Neumann equation for
joint density matrix. At $v_k(t)\neq 0\,$, the trace
of $R(t)$ in $D\otimes B$ represents [1,5] the characteristic
functional of the observables, $\Xi (t,v)$:
\[
\text{Tr}_D\text{Tr}_B\,R=\text{Tr}_D\text{Tr}_B\,\,
\overrightarrow{\exp }\left [\frac 12 \int v(t^{\prime })
J(t^{\prime })dt^{\prime }\right]\times
\]
\begin{equation}
\times\overleftarrow{\exp } \left[
\frac 12\int v(t^{\prime })
J(t^{\prime })dt^{\prime }\right]R_{in}=\Xi (t,v)
\end{equation}
Here $R_{in}=$ $R(-\infty)$,
the presence of time argument in $J_k(t)$
means that these operators are treated in the Heisenberg
picture, left-hand (right-hand) oriented arrow
indicates chronological (anti-chronological) time ordering,
and the integrals are taken over the region $t^{\prime}<t$
(with $t$ being present time moment).
Let us replace the operators $B(t) $ in Hamiltonian (2)
by commutative (similar to $C$-numbers)
variables $\xi(t) =$ $x(t)+iy(t)/2$
or $\eta(t) =$ $x(t)-iy(t)/2$
and, instead of Eq.3, consider the equation
\begin{equation}
\dot{\rho}(t)=v(t)J\circ \rho(t) +L(t)\rho (t)\,,
\end{equation}
\begin{equation}
L(t)\rho \equiv i\left
\{\rho H(\eta ,t)-H(\xi ,t)\rho \right \}=
\end{equation}
\[
=\sum_j y_j(t)D_j\circ \rho +i[\rho ,H(x,t)]\,,
\]
treating all the $\eta(t)$, $\xi(t) $, $x(t)$ and $y(t)$
as random processes which reflect an influence
by the thermostat.
Assume that $R_{in}$ has the factorized
form:  $R_{in}=$ $\rho(-\infty)\rho _b$,
where $\rho _b=$ $\text{Tr}_D\,R(-\infty)$
is statistical operator of the thermostat.
All of $x(t)$ and $y(t)$ are defined
if their characteristic functional
is defined. We make it as follows:
\begin{equation}
\left\langle \exp \,
\int [g(t)x(t)+f(t)y(t)]dt\right\rangle=
\end{equation}
\[
=\text{Tr}_B\,\,\overrightarrow{\exp }\left\{
\frac 12 \int g(t)B(t,f)dt\right\}\times
\]
\[
\times\overleftarrow{\exp }
\left\{ \frac 12
\int g(t)B(t,f)dt\right\}\,\rho _b
\]
(the subscripts are omitted for brevity),
where the operators $B_j(t,f)$
are defined by means of the equations
\begin{equation}
B_j(t,f)= U^{+}(t,f)B_jU(t,f)\,,
\end{equation}
\[
\dot{U}(t,f)=-i\{H_b+\sum_j f_j(t)B_j \}U(t,f)
\]
Clearly, the time-dependent operators $B_j(t,f)$
characterize non-autonomous behaviour of the thermostat.
On the left side of Eq.7, $g(t)$ and $f(t)$
are test functions for the random sources $x(t)$ and $y(t)$,
respectively. On the right side,
according to Eqs.7 and 8, $g(t)$ are test functions for
the thermostat observables $B(t)$ while $f_j(t)$ are
classical ($C$-number valued) forces which perturb
the thermostat being conjugated with $B_j(t)$
(in the sense of non-equilibrium thermodynamics).
Note that the relation (7)
can be rewritten in the form [1]:
\[
\left\langle \exp \,\int
\sum_j[g_j(t)x_j(t)+f_j(t)y_j(t)]dt\right\rangle _b\equiv
\]
\[
\equiv \text{Tr}_B\,\,
\overrightarrow{\exp }\left\{
\int \sum_j\left[
\frac{ g_j(t)}2+if_j(t)\right] B_j(t)dt\right\}\times
\]
\[
\times\overleftarrow{\exp }\left\{ \int
\sum_j\left[ \frac{g_j(t)}2-if_j(t)\right] B_j(t)dt\right\}
\,\rho _b\,,
\]
demonstrating formally equal
rights of $x(t)$ and $y(t)\,$ on the right-hand side
of Eq.7.

As it was shown in [1], if statistics of
commutative random sources
in Eqs.5 and 6 is determined by Eqs.7 and 8 then
the sources exactly imitate real quantum thermostat.
Concretely,
\begin{equation}
\left\langle\rho(t)\right\rangle=\text{Tr}_B\,R(t)\,,
\end{equation}
\begin{equation}
\text{Tr}_D\,\left\langle \rho (t)\right\rangle =\Xi(t,v)\,,
\end{equation}
where the angle brackets $\left\langle ...\right\rangle $
denote statistical averaging with respect to $x(t)$ and $y(t) $
(or to $\eta(t) $ and $\xi(t) $) in accordance with (7-8).

\,\,\,

{\bf 2. Properties of the random sources.}

The pay for the desired replacement is the doubling of
random sources: instead of each real observable
(Hermitian operator) $B_j(t)$ we obtain the pair of
either real variables ($x_j(t)$ and $y_j(t)$) or
mutually conjugated complex
variables ($\eta_j\,$ and $\xi_j= $ $\eta_j^{*}$).
From Eqs.5-6 it is seen that $x_j(t)$
play direct role of random forces (potentials),
while $y_j(t)$ stay in positions of test functions
which correspond to observing DS by thermostat
(in analogy with $v_k(t)$ in Eq.3).
In absence of $y_j(t)$, under random pump
introduced by $x_j(t)$, both the energy and entropy
of DS would grow as much as possible.
However, like in any measurement process,
influence of $y_j(t)$ introducing observation by
thermostat to the ``stochastic Liouville
operator'' (6) destroys unitarity of evolution of
the statistical operator $\rho (t)$
and decreases phase volume (entropy) of DS.
At the same time $y_j(t)$ are responsible for
the energy outflow back to thermostat, i.e. for
dissipation, and hence for non-uniform (thermal)
DS energy distribution over states of DS.
It should be emphasized that according to (9-10)
on the average the unitarity is ensured.

Due to Eqs.7-8,
\begin{equation}
\left\langle \prod_jx(t_j)\prod_my(t_m^{\prime })
\right\rangle =
\end{equation}
\[
=\left[
\prod_m\frac \delta {\delta f(t_m^{\prime })}\left\langle
\prod_jB(t_j,f)\right\rangle \right] _{f=0}
\]
Here from it is clear
that $\left\langle y(t) \right\rangle$  $=0 $,
 $\,\left\langle y(t)y(t^{\prime }) \right\rangle$  $=0 $,
and all the higher self-correlators of $y(t) $
also are equal to zero. But the cross-correlators
between $y(t) $ and $x(t) $ differ from zero
representing the response of thermostat to its
perturbation by DS.
Thus, $y(t) $ are not $C$-numbers in literal sense.
But the singularity of statistical
properties of $y(t)$ only facilitates factual calculations.
Importantly, $y(t) $ can possess correlations with more
late values $x(t^{\prime}>t) $ only, in correspondence
with the causality principle.

In case of Gaussian equilibrium (``bosonic'') thermostat,
all the functions (11) can be expressed through
pair correlators
\begin{equation}
\left\langle x_j(\tau)x_m(0) \right\rangle =
\int_0^\infty \cos (\omega \tau )S_{jm}(\omega )
\frac {d\omega}{\pi}  \,,
\end{equation}
\[
\left\langle x_j(\tau)y_m(0) \right\rangle
=2\vartheta (\tau )\int_0^\infty
\sin (\omega \tau )\times
\]
\[
\times\tanh \left(
\frac {\omega}{2T} \right) S_{jm}(\omega )
\frac {d\omega}{\pi}
\]
Here $T $ is temperature of thermostat, $\vartheta (\tau )$
is Heavyside function, and $S_{jm}(\omega )$
is a non-negatively defined spectral matrix.
The formulas (12) show that neglecting the $y(t)$-component
of random sources in (5-6) in essence would be
equivalent to infinity of thermostat temperature.

In Eqs.9-10, only the average value of the
``stochastic density matrix of DS'', $\rho(t) $,does appear.
Naturally, its higher statistical moments
\begin{equation}
\left\langle \rho(t)\otimes ...
\otimes\rho (t)\right\rangle
\end{equation}
describe several copies of DS interacting with the same
thermostat. The undistinguishness of the copies require
to concretize their quantum statistics (second quantization
rule). In [1] we considered how the case of Fermi
statistics reduces to analysing the moments (13), thus
replacing second quantization in a ``many-particle
problem'' by equivalent ``second randomization'' in
one-particle problem. In this approach, an additional
direct (e.g. Coulombian) interaction between samples of DS
can be included as interaction through the second
(Gaussian) thermostat, i.e. additional pairs of
stochastic sources $x(t) $, $y(t) $.

\,\,\,

{\bf 3. Two-level system in Gaussian thermostat.}

Let us apply the above formalism to two-level
system (TLS) interacting with Gaussian thermal bath.
We can put on
\begin{equation}
J=\left(
\begin{array}{cc}
1 & 0 \\
0 & -1
\end{array}
\right) \,,\,\,
D=\left(
\begin{array}{cc}
0 & 1 \\
1 & 0
\end{array}
\right)\,,
\end{equation}
and $H_d(t)=u(t)J/2\,$.
Here $u(t)$ is energy difference between two states,
possibly depending on time because of external
perturbation of TLS. The operator $D $ performs contact
with thermostat which forces random switchings
of TLS state. The operator $J$ corresponds to observation
of TLS state, for instance, spin orientation, or
alternate velocity of quantum particle in the simplest
model of Brownian motion. We use this model to illustrate
role of the $y(t)$-component of random source, i.e.
``dissipative'' term $y(t)D\circ\rho(t) $ in Eqs.5-6.

First, let both states of free TLS have equal
energies: $u(t)\equiv 0$. Then the Eqs.5-6 can be
integrated up to rather visual expressions.
In particular, for the trace
\[
\theta (t,v)\equiv \text{Tr}_D\,\rho (t)=
\rho_{11}+\rho_{22}
\]
one can derive the recurrent equation
\begin{equation}
\theta (t,v)=Y(t,-\infty)+
\end{equation}
\[
+\int Y(t,t_1)v(t_1)X(t_1,t_2)v(t_2)
\theta (t_2,v)dt_1dt_2\,,
\]
\[
X(t_1,t_2)\equiv \cos \left [2
\int_{t_2}^{t_1}x(t^{\prime})dt^{\prime} \right]\,,
\]
\[
Y(t_1,t_2)\equiv \cosh \left [
\int_{t_2}^{t_1}y(t^{\prime})dt^{\prime} \right]\,,
\]
where $t>t_1>t_2$,
and $\Xi (t,v)=$ $\left\langle \theta (t,v)\right\rangle $.

Consider the correlation function of spin fluctuations
(or velocity fluctuations if speak about Brownian
particle, or so on),
 $K(\tau)=$ $\left\langle J(\tau)J(0) \right\rangle $,
and the diffusivity
 $\Delta =$ $\int_0^{\infty}K(\tau )d\tau $.
First of all the case is curious when the noise of
thermostat is ``white'', i.e. $S(\omega)=const$ in (12).
To be precise, it should be underlined that both the
correlators (12) can not simultaneously turn
into $\delta$-functions, and in this sense quantum
noise never can be white. After
expanding Eq.15 into series, averaging it with
taking into account Eqs.12 and then differentiating
by the test function, one obtains
\begin{equation}
K(\tau)=\text{e}^{-2S\tau} \cos \left \{
\frac {2S}{\pi}
\int_0^{\tau} \ln
\tanh \left(\frac{\pi T\tau^{\prime}}{2}
\right)d\tau^{\prime}
\right \}
\end{equation}
Under increase of the order of
magnitude of the dimensionless noise intensity, $S/T$,
initially monotonous relaxation changes to more and more
oscillating relaxation. From the formal point of view,
this is just the effect of the $y(t)$-component, i.e.
inverse influence by DS onto thermostat
(indeed, in absence of $y(t)$ the correlation would
be purely exponential independently on $S/T$).

Howevere, practically the opposite approximation can
occur more appropriate:
\begin{equation}
S(\omega)=S/[1+(\tau_0\omega)^2]\,,\,\,\tau_0T>>1\,,
\end{equation}
when sufficiently large correlation time of the
thermostat noise (characteristic time of thermostat
response), $\tau_0$, excludes high frequencies.
Then it is convenient to introduce the coupling
energy $\epsilon $ by
means of $S(0)=$ $2\epsilon^2\tau_0\,$.
Calculating the correlation function under assumptions
(17), we find again that at weak coupling ($\epsilon/T<1$)
monotonous (nearly exponential) relaxation takes place,
while at $\epsilon/T>1$ (strong coupling)
\begin{equation}
K(\tau)\approx \exp(-2\epsilon^2\tau^2)
\cos (\epsilon^2\tau/T)\,,
\end{equation}
that is again the oscillations arise and
then become multiplied.

Clearly, oscillations of relaxation lead to an
excess decrease of the diffusivity $\Delta $
(spectral power density of $J(t)$ at zero frequency)
as compared with its value $\Delta_0 $ which would
realize in absence of $y(t) $.
In the case of Lorenz noise with $T\tau_0$ $>1$,
the diffusivity can become extremely
small resulting in spatial localization of
Brownian particle.

The non-monotonous relaxation means that the TLS response
to periodic external perturbation can acquire ``resonant''
character having maximum at some non-zero frequency of
the perturbation (or, if the frequency is fixed, maximum
at some non-zero level of thermostat noise). Such a kind
of phenomena is known as ``stochastic resonance'' (see,
for example, the review [2]).

We would like to emphasize also that oscillatory
relaxation represents non-Marcovian effect which
could not be adequately considered by
the theory of Marcovian quantum dynamical
semi-groups, i.e. time-local kinetic equations
(this theory takes beginning from the
classical work [3]). In terms
of this theory, we considered ``stochastic dilation''
of a semi-group which is as much non-Marcovian
as strongly the correlator
 $\left\langle x(\tau)y(0)\right\rangle $
determining oscillations in (16) and (18)
differs from $\delta$-function.

\,\,\,

{\bf 4. Stochastic linear response of TLS to
non-stochastic perturbation.}

Because of the symmetry (degeneration) of states of TLS,
the trace (15) contains even powers of $v(t)$ only.
If $u(t)\neq 0$ then this symmerty breaks down.
Define integral operators $\hat{I}$, $\hat{C}$
and $\hat{S}$ by the formulas
\[
\hat{I}f(t)=\int_{-\infty}^tf(t^{\prime})dt^{\prime}\,,
\]
\[
\hat{C},\hat{S}\,f(t)=\int_{-\infty}^t
\cos,\sin
\left\{\int_{t^{\prime}}^t
u(t^{\prime\prime})dt^{\prime\prime}
\right\}f(t^{\prime})dt^{\prime}\,,
\]
At $u(t)\neq 0$ instead of (15) the following
equation can be derived:
\begin{equation}
\theta=1+\hat{I}y\hat{C}y\theta+
\end{equation}
\[
+\hat{I}(v+2y\hat{S}x)
[1+4\hat{I}x\hat{C}x]^{-1}
\hat{I}(v+2x\hat{S}y)\theta
\]
Now, in the expansion of $\theta$ into
power series of $v(t)$,
\[
\theta(t,v)=\theta_0(t)+\int_{-\infty}^t
\widetilde{J}(t_o)v(t_o)dt_o+
\]
\[
+\frac {1}{2}\int_{-\infty}^t\int_{-\infty}^t
\widetilde{K}(t_1,t_2)v(t_1)v(t_2)dt_1dt_2+\dots\,,
\]
the first order term is present (as well as other odd
terms). For simplicity, consider this contribution
in the ``hot thermostat
limit'': $S/T<<1\,$, $T\tau_0>>1\,$,
when in accordance with (12) any multiplier $y(t)$
contributes excess order of smallness.
Besides, assume the perturbation $u(t)$ is infinitely weak.
Then, saving lowest powers of $y(t)$ and $u(t)$ only,
we can write
\begin{equation}
\widetilde{J}(t_o)=J_{+}(t_o)+J_{-}(t_o)\,,
\end{equation}
\[
J_{+}=\int_{-\infty}^{t_o}
\sin \left\{2\int_{t_1}^{t_o}x(t^{\prime})dt^{\prime}
\right\}\times
\]
\[
\times u(t_1)\int_{-\infty}^{t_1}y(t_2)dt_2dt_1\,,
\]
\[
J_{-}=\int_{t_o}^t
\sin \left\{2\int_{t_o}^{t_1}x(t^{\prime})dt^{\prime}
\right\}u(t_1)\int_{t_1}^ty(t_2)dt_2dt_1
\]
Of course, here the factual observation time $t_o$
is more early than formal running time $t$.
Under this condition, $t $
drops out from any result of statistical averaging,
therefore it may be convenient to set $t=\infty$.

Obviously, $\widetilde{J}(t_o)$ disappears if we
neglect $y(t)$-component of the bath noise.
Consequently, this component (in company with $x(t)$)
is responsible for non-zero response of DS to
external perturbations.
The integral operators which act onto $u(t)$
in the expressions $J_{+}$ and $J_{-}$
are nothing but random linear response functions
(random susceptibility, or mobility, or so on),
in other words, these operators represent
multiplicative (parametric) noise.
What is hidden beyond it is the randomness of
quantum probabilities of quantum transitions,
i.e. dependence of the probabilities on thermostat
noise [4].

Formulas (20) clearly highlight that
the response functions in their non-averaged form
has no decay and can connect past and future events
regardless of how far they are distanced in time.

\,\,\,

{\bf 5. Two TLS in shared thermostat.}

Consider two identical TLS contacting with one and
the same thermostat. Suppose that the systems are formally
distinguishable and that they give additive contributions
to the variable under observation. Then characteristic
functional of the summary observable is presented by
\begin{equation}
\Xi(t,v)=\left\langle \theta^2(t,v)\right\rangle\,,
\end{equation}
instead of (10).
For its correlation function,
to be designated as $K_2$, in the hot thermostat limit
the Eq.21 yields
\[
K_2(t_1,t_2)=2\{
\left\langle \widetilde{K}(t_1,t_2)\right\rangle+
\left\langle J_{+}(t_1)J_{-}(t_2)\right\rangle+
\]
\begin{equation}
+\left\langle J_{+}(t_1)J_{+}(t_2)\right\rangle-
2\left\langle J_{+}(t_1)\right\rangle
\left\langle J_{+}(t_2)\right\rangle\}\,,
\end{equation}
where $t_1>t_2$ is mentioned.

We do not discuss the first term on the right-hand
side of Eq.22. What is for the next terms,
they describe ``excess noise'' of summary observable
which is proportional to squared perturbation $u^2$
and comes just from the fluctuations in random linear
responses (similar in both TLS and determined
by Eqs.20). If $t_1-t_2$ $>>\tau_c$,
where $\tau_c$ is correlation time of the fast component
of the $J(t)$ fluctuations (that is decay time of
equilibrium correlation function $K(\tau)$ at $u(t)=0$),
then a contribution from the cross-correlator
of $J_{+}$ and $J_{-}$
does survive only, and further it does not decay at all.
A rough formal estimate of these long-living fluctuations
by order of magnitude gives
\[
|\left\langle J_{+}(\infty)J_{-}(-\infty)\right\rangle|
\sim (\tau_c^2/\tau_0^2)
\left\langle \widetilde{J}\right\rangle^2
\]
Here the ratio $\tau_c/\tau_0$ crucially depends on the
parameter $\epsilon\tau_0$ and can be much greater
than unit (at $\epsilon\tau_0$ $<<1$) as well as
much smaller than unit (at $\epsilon\tau_0$ $>>1$).

Interestingly, the expression $J_{-}(t_o)$
in (20) corresponds to
``reaction of the past on the future'' because
it includes future values of all the variables only
with respect to the observation time.
With no doubts, any correlator in which time argument
of $J_{-}$ is most late should turn into zero,
otherwise it would have no physical sense.
However, non-zero correlations between $J_{-}$
and more late $J_{+}$ does not contradict to causality
principle.

The non-decaying tails of such correlations
in two-TLS system are quite similar to
properties of fluctuations of quantum transport
probabilities in many-electron
system considered in [4].
Their physical origin is the coherence (unitarity and time
reversibility) of joint evolution of DS and thermostat:
the coherence can be interrupted by external observation
acts but conserves between them.
What is important, unlike [4], in the present example
formal analysis of non-decaying correlations
is not restricted by any time frames.

\,\,\,

 ---------------------------------------------------

\,\,\,

{\bf 6. Appendix. Generalized stochastic
representation.}

The text above almost coinsides with the
paper in Russian [5] (to be translated into English
in JETP Letters). Since the proof of the basic
relations (9),(10) is omitted in [5] (see [1] for it),
below we compensate this by proving generalized
stochastic representation which includes the above
one as a particular case.

Let the joint evolution (Liouville)
operator, $L\,$, of combined
system ``DS plus thermostat'' has the
bilinear form as follows:
\begin{equation}
L=L_D+L_B+\sum_j \Lambda _{Dj}\Lambda _{Bj}\,,
\end{equation}
where $L_D$ and $L_B$ are evolution (Liouville)
operators of autonomous DS and thermostat, respectively,
and $\Lambda_{Dj}$ and $\Lambda_{Bj}$ are
some superoperators acting in $D\,$ and $B\,$,
respectively (to be precise, they act in the spaces
of linear operators defined in $D\,$ and $B\,$).
For instance, the superoperators $\Lambda_{Dj}$
and $\Lambda_{Bj}$ can be Liouville
operators (commutators) or multiplication
operators (Jordan products). For shorteness,
assume that the terms $\sum v_k(t)J_k$
corresponding to external watching for DS
are included into $L_D$.

Notice, that, from one hand,
if $\alpha(t) $ is any random process
with characteristic functional
\begin{equation}
\Psi\{f\}=\left\langle \exp\left[\int
f(t)\alpha(t)dt\right]\right \rangle
\end{equation}
(angle brackets mean averaging over $\alpha$'s
probability distribution),
then the average of any functional $\Phi\{\alpha\}$
can be formally represented by
\begin{equation}
\left\langle \Phi\{\alpha\}\right \rangle=
\Psi\left\{\frac {\delta}{\delta \alpha}
\right\}\Phi\{\alpha\}\,_{\alpha=0}
\end{equation}
Similar relation holds also for a set
of random processes $\alpha_j(t)$.

From another hand,
since all of $L_D$ and $\Lambda_{Dj}$ commute
with all of $L_B$ and $\Lambda_{Bj}$,
we can write:
\begin{equation}
\overleftarrow{\exp }\left\{\int\left[
L_D+L_B+\sum \Lambda_{Dj}\Lambda_{Bj}
\right]dt \right\}=
\end{equation}
\[
=\overleftarrow{\exp }\left\{\int\left[
L_B+\sum \Lambda_{Bj}\frac
{\delta}{\delta\alpha_j(t)}\right]dt \right\}\times
\]
\[
\times\overleftarrow{\exp }\left\{\int\left[
L_D+\sum \alpha_j(t)\Lambda_{Dj}
\right]dt \right\}\,_{\alpha(t)=0}
\]
Here mutual chronological ordering of the two chronological
exponents in their product is taken in mind.
This is merely the consequence from chronological
ordering and besides from the formal identity
\[
\exp(O_1O_2)=\exp\left(O_1\frac {\partial}
{\partial \alpha}\right)
\exp(\alpha O_2)\mid_{\alpha=0}\,\,,
\]
where $O_1$ and $O_2$ are arbitrary mutually
commuting objects: $[O_1,O_2]=0\,$.

Suppose that initial statistical operator $R_{in}$
factorizes, $R_{in}=$ $\rho(-\infty)$ $\rho_b\,$
where $\rho $ stands for DS density matrix.
Then, comparing formulas (25) and (26),
we conclude, firstly, that
from the point of view of DS its joint
evolution together with thermostat is equivalent
to its individual evolution but influenced by
the set of random sources $\alpha_j(t)$ and
governed by stochastic evolution operator
as follows:
\begin{equation}
\dot{\rho}=\left\{v(t)J\circ\rho +L_D+
\sum \alpha_j(t)\Lambda_{Dj}\right\}\rho
\end{equation}
(we extracted external DS observation
back from $L_D$).
Secondly, in analogy with (24) and (25),
characteristical functional of these sources
is determined by
\begin{equation}
\left\langle\exp\left[\int\sum
f_j(t)\alpha_j(t)dt\right]\right\rangle=
\end{equation}
\[
=\text{Tr}_B\,\overleftarrow{\exp}\left\{\int
\left[ L_B+\sum f_j(t)\Lambda_{Bj}
\right]dt\right\}\,\rho_b
\]

Formulas (27) and (28) give the stochastic
representation of the joint evolution.
In fact, it is obtained by nothing but a variant
of the Stratonovich transformation.
Dependently on concrete contents of
operators $\Lambda _{Dj}$ and $\Lambda _{Bj}$
the random sources $\alpha_j(t)\,$ can
behave either similar to usual classical
random processes (like $x(t)$ above) or in a
singular manner (like $y(t)$ above) or in some
mixed fashion. Of course, in general all the
sources are mutually correlated.

In particular, the above considered case
of the bilinear interaction Hamiltonian (1)
can be reduced to the representation (27)-(28)
or derived from it if take into account the
identity
\begin{equation}
[R,DB]=[R,D]\circ B+[R,B]\circ D
\end{equation}
which is valid for any pair of mutually
commuting $D$ and $B\,$, at $[D,B]=0\,$. Hence,
every pair $D_j$ and $B_j$ from (1)
produces two terms in (23). Clearly, in the
corresponding pair of $\alpha$'s one conjugated
with the first term on the right-hand side of (29)
plays the role of $x(t)$ (see Sec.1), while
another (conjugated with second term) plays
the role of $y(t)$.

It is useful to demonstrate a simple case
which does not reduce to bilinear interaction (1).
Concretely, let the joint evolution operator $L $
is defined by
\begin{equation}
LR=(L_D+L_B)R+iD^{\prime}\circ [R,B]
+iB^{\prime}\circ [R,D]\,,
\end{equation}
where $D^{\prime}$ differs from $D$,
$\,B^{\prime}$ differs from $B$,
and $L_{D,B}R=$ $i[R,H_{d,b}]$.
Then, instead of the representation expressed
by Eqs.5-8, we arrive to the representation
\begin{equation}
\dot{\rho}=v(t)J\circ\rho +y(t)D^{\prime}\circ\rho+
i[\rho,H_d+x(t)D]\,\,,
\end{equation}
\begin{equation}
\left\langle \exp \,
\int [g(t)x(t)+f(t)y(t)]dt\right\rangle=
\end{equation}
\[
=\text{Tr}_B\,\,\overrightarrow{\exp }\left\{
\frac 12 \int g(t)B^{\prime}(t,f)dt\right\}\times
\]
\[
\times\overleftarrow{\exp }
\left\{ \frac 12
\int g(t)B^{\prime}(t,f)dt\right\}\,\rho _b\,\,,
\]
\begin{equation}
B^{\prime}(t,f)= U^{+}(t,f)B^{\prime}U(t,f)\,\,,
\end{equation}
\[
\dot{U}(t,f)=-i\{H_b+f(t)B\}U(t,f)
\]
Now the two random sources involve not two
but four operators $D$, $D^{\prime}$, $B$
and $B^{\prime}$.
Nevertheless, again it is possible to
separate analysis of thermostat, under
given classical perturbation, and analysis of DS under
influence by commutative stochastic sources
with given statistics (and again statistical properties
of a number $N$ of DS copies interacting with the same
thermostat can be obtained from the 
moments $<($Tr$_B\,\rho)^N>$,
if the copies are distinguishable and undergo
Boltzmann statistics, or as it was
considered in [1], if the copies undergo Fermian
second quantization statistics).
Concrete application of this formal example
will be done elsewhere.

\,\,\,

{\bf References}

\,\,\,

1. Yu.\,E. Kuzovlev, cond-mat/0102171.

2. L. Cammaitoni, P. Hanngi, P. Jung et al.,
 Rev. Mod. Phys. {\bf 70}, 223 (1998).

3. G. Lindblad, Commum. Math. Phys. {\bf 48}, 119 (1976).

4. Yu.\,E. Kuzovlev, Yu.\,V. Medvedev, A.\,M. Grishin,
JETP Letters, {\bf 72}, 574 (2000).

5. Yu.\,E. Kuzovlev, Pis'ma v ZhETF, {\bf 78}, 103 (2003).


\end{document}